\documentstyle[12pt,epsf,epsfig]{article}

\begin{document}

\title{A new perspective on the analysis of
helix-helix packing preferences in globular proteins}
\author{A. Trovato and F. Seno \\
 INFM-Dipartimento di Fisica `G. Galilei',
Universit\`a di Padova, \\ Via Marzolo 8, 35131  Padova, Italy }

\maketitle

\newpage

\begin{abstract}

For many years it had been believed that steric compatibility of helix
interfaces could be the source of the observed preference for
particular angles between neighbouring helices as emerging from
statistical analysis of protein databanks. Several elegant models
describing how side chains on helices can interdigitate without steric
clashes were able to account quite reasonably for the observed
distributions.  However, it was later recognized (Bowie, 1997 and
Walther, 1998) that the ``bare'' measured angle distribution should be
corrected to avoid statistical bias.  Disappointingly, the rescaled
distributions  dramatically lost their similarity with
theoretical predictions casting many doubts on the validity of the
geometrical assumptions and models. In this report we elucidate a few
points concerning the proper choice of the random reference
distribution.  In particular we show the existence of crucial corrections
due to the correct implementation of the approach used to discriminate
whether two helices are in contact or not and to measure their
relative orientations.  By using this new rescaling, the ``true''
packing angle preferences are well described, even more than with the
original ``bare'' distribution, by regular packing models.

\end{abstract}

\newpage

\section*{Introduction}

The issue of the pairwise packing between helices in proteins was
addressed soon after helical structures had been suggested.  A number
of models were developed, mostly devoted to surface
complementarities upon packing. The ``knobs into holes'' model, first
introduced by Crick \cite{Crick53} and elaborated by Richmond and
Richards \cite{RR78}, aimed to find the best steric fit between regular
helices. Chothia, Levitt and Richardson
\cite{Chothia77,Chothia81} recognized the importance of ``ridges''
 and ``grooves'' formed by residues with different sequential
distances (``ridges into grooves'' model).  Efimov \cite{Ef79} tried to
relate the packing angle between the two helices with the preferred
rotational states of the side-chains along them.
A comprehensive analysis was eventually carried out
by Walther et. al. \cite{Walther1}, by modeling helix packing as the
superposition of the two regular lattices that result from unrolling the
helix cylinders onto a plane and contain points representing each
residue. The six ``preferred'' angles predicted by this last model
are consistent with earlier results and with the histograms
of the experimentally observed packing angles (see Fig. 7 of Ref.
\cite{Walther1}).
The agreement between theoretical modeling and experimental data
was remarkable, although not perfect (see Ref. \cite{Walther1} for
a more detailed discussion).

The success of steric models in providing an explanation for the most
prevalent packing angles was however put under discussion after the
observation made by Bowie \cite{Bowie97} that statistical corrections
must be applied to the values collected from experimentally determined
structures before true interaxial angle
preferences can be revealed. Indeed, the helix-helix packing is
defined to occur only when:

\begin{itemize}
\item{[C1]}  the segment of closest approach (SCA), of length $d_R$, between
the two finite helix axes is shorter than a prefixed threshold $d_c$;

\item{[C2]} this segment intersects both helix axes at a perpendicular angle.

\end{itemize}

In Ref. \cite{Walther1}, two helices were considered to be in contact
if they were satisfying both conditions $C1$ and $C2$.  Note that
Walther {\it et al.}\/ \cite{Walther1} selected interacting helical
pairs from their database of native protein conformations, by
considering distances between all possible inter-helical heavy atom
pairs. We used a different selection procedure involving distances
between inter-helical $C_{\alpha}$ atom pairs (see Methods for
details). None of these particular choices really affects our
geometrical analysis, and condition C1, as defined above,
is just the simplest way of defining an equivalent distance
constraint.

Condition C2 is needed to ensure face-to-face packing of the two
helices, justifying thus the use of theoretical modeling based on the
steric interdigitation of the helices.

If the SCA is coincident with the global segment of closest approach
(GSCA), of length $d$, between the two straight lines which are
obtained by indefinitely prolonging the helix axes, condition C2 is
then automatically satisfied.  In such a situation, which corresponds
to effectively dealing with helices of infinite length (see Fig. 1a),
the reference probability distribution of interaxial angles
$P(\Omega)$ is simply the spherical-polar distribution between any two
random vectors, namely $P(\Omega)\sim\sin{\Omega}$ \cite{Bowie97}.

Walther and co-workers \cite{Walther2} realized that the finiteness of
helix axes is crucial in modifying the angular dependence of $P(\Omega)$,
because requiring that the SCA intersects both axes at a perpendicular
angle introduces new
restrictions depending on the packing angle $\Omega$. This can be understood
by looking at Fig. 2a, where it is shown that, fixing one helix position,
and assuming the GSCA to be orthogonal to the page,
the second finite helix axis may then be placed
only within a plane parallel to the paper plane and such that its
starting point lies inside the dark shaded parallelogram $A$.

The probability of placing the starting point of the second helix in
A is proportional to its area and therefore to
$\sin\Omega$. Consequently the probability $P({\Omega})$ for
selecting a particular packing angle is proportional to the product of
the spherical-polar contribution and of the
$\sin \Omega$ effect due to the finite length condition:

\begin{equation}
P (\Omega) \sim \sin^2\Omega
\end{equation}

With this new random reference distribution the actual angular
propensities need to be reconsidered. As is clear from Fig. 2 of
Ref. \cite{Walther2}, the normalized frequencies reveal a prominent
representation of packing angles near $0^{\circ}$ and $180^{\circ}$
(not expected by
theoretical models) whereas the predicted optimal steric packing angles
manifest themselves, at most as shoulders in the distribution of
propensities. The predictivity of geometrical models is cast under serious
doubts after this analysis.

Our first observation is that the random distribution
$P(\Omega)\sim\sin^2\Omega$
is correct only and only if the condition of mutual
perpendicularity between the SCA and the two
axes is strictly fulfilled.
But is this the real situation when statistical histograms are derived?

As a matter of fact, condition $C2$ was relaxed by Walther {\it et al.}\/,
by admitting a small tolerance: a total deviation $\tau=\tau_1+\tau_2$
($\tau_1$ and $\tau_2$ are the complementary angles to the
angles $\theta_1$ and $\theta_2$ formed between the SCA
and helix 1 and helix 2, respectively (see Fig. 1b)) was accepted up to
a threshold  $\tau_{\rm max}$ ($\tau_{\rm max}=5^{\circ}$ in \cite{Walther1}
and $\tau_{\rm max}=1^{\circ}$ in \cite{Walther2}).

At first, one might wonder why such a threshold needs to be used since
it does not effectively increase the number of data contributing to
the histograms. However, we believe that this choice is indeed
necessary because of the ambiguity in the definition of the axis
direction in natural helices, which introduces an intrinsic
uncertainty in the computation of the $\Omega,\theta,\tau$ angles (see
Methods for a detailed explanation of how the axis is reconstructed in
the typical case of a non-ideal bent helix and for an estimation of
such uncertainty $\overline{\Delta\Omega}$).  A relaxation of
condition $C2$ is thus crucial, if we want to analyse correctly data
extracted from the database of real native protein structures.

We will show in this paper, by means of partly semi-analytical
geometrical arguments and eventually numerical simulations, that such
threshold effect does drastically change the random reference
distribution.  Having measured helix-helix packing angles from a set
of $600$ proteins representative of the PDB native structures, we will
then reanalyse the packing angle distribution after its proper
rescaling with the newly found reference distribution.

\section*{Methods}

\subsection*{Databank and helix pair selection}

We employed the same ensemble of 600 proteins considered by Chang
et. al \cite{Iksoo}, which consisted of sequences varying in length
from 44 to 1017, with low sequence homology and covering many
different three-dimensional folds according to the Structural
Classification of Proteins (SCOP) scheme \cite{Murzin}.  The
structures were monomeric and determined using x-ray crystallography.
We collected $4397$ helices each with at least four consecutive
residues classified as helical in the PDB files.  The average number
of residues of these helices was $11.5$.  Two helices were defined to
be in close contact, if at least one interhelical contact between
$C^{\alpha}$ atoms was present, with a maximal threshold distance of
$5.5 \AA$ (analogous to condition $C1$ in the text). Only helix pairs
separated in sequence by at least $40$ intervening residues were
considered, to get rid of possible correlations induced by short
loops. The resulting data set consisted of $1460$ closely packed helix
pairs.

\subsection*{Helix axis reconstruction}

The reconstruction of the helix axis from the coordinates of the
$C^{\alpha}$ atoms of the corresponding residues is a critical
step in the determination of packing angle preferences.
Since real helices can be bent, we adopted the procedure described
by Walther {\it et al.}\/ in ref. \cite{Walther1}, in which a local axis
is associated to every consecutive residue pair along the helix.
The overall axis is thus a broken line consisting of short segments.

A good starting approximation for the local axis ${\bf a}_i$, based on the
the four $C^{\alpha}$ atom positions ${\bf r}_{i-1}$, ${\bf r}_i$,
${\bf r}_{i+1}$, ${\bf r}_{i+2}$, was
introduced by Chothia {\it et al.}\/ \cite{Chothia81}.
We employed a sligthly modified definition, where we first define
the normalized bond vectors
${\bf b}_i = \left({\bf r}_{i+1} - {\bf r}_i\right)/
\left|{\bf r}_{i+1} - {\bf r}_i\right|$.
In this way, the set of three orthonormal vectors ${\bf t}_i$, ${\bf v}_i$,
${\bf u}_i$, the natural reference system associated with the
$C^{\alpha}$ atom trace, can be defined without the distortions due
to the fluctuations of the bond length between consecutive $C^{\alpha}$ atoms
in the following way \cite{Rey92}:
${\bf t}_i= \left({\bf b}_i+{\bf b}_{i-1}\right)/
\left|{\bf b}_i+{\bf b}_{i-1}\right|$,
${\bf v}_i={\bf b}_i\times{\bf b}_{i-1}/
\left|{\bf b}_i\times{\bf b}_{i-1}\right|$,
${\bf u}_i = {\bf t}_i\times {\bf v}_i$.
If the $C^{\alpha}$ atom trace followed a perfect ideal helix, the vector
${\bf a}_i = {\bf u}_i\times{\bf u}_{i+1}$ would be parallel to the
helix axis. We then initially determined the local axis between residues
$i$, $i+1$, as the vector parallel to ${\bf a}_i$, of length $1.45 \AA$,
which has the geometric centre of the closest four consecutive $C^{\alpha}$
atom positions, ${\bf r}_{i-1}$, ${\bf r}_i$, ${\bf r}_{i+1}$,
${\bf r}_{i+2}$, as its midpoint. The two local axes at the helix termini
were obtained by simply prolonging the neighbouring ones.

We then applied a smoothing procedure similarly to Walther {\it et al.}\/
\cite{Walther1}. The `new' local axes ${\bf a}_i$ were obtained by
averaging the direction of the closest three `old' ones,
${\bf a}_{i-1}$, ${\bf a}_i$, ${\bf a}_{i+1}$, while conserving its
midpoints.
For the two local axes at the helix termini we averaged over the direction
of the closest two available ones. To ensure the continuity of the overall
axis, the inner hinges of the broken axial line were computed as the midpoints
between extremities of consecutive local vectors obtained from
the running average.
After repeating the whole procedure twice, the standard deviation
of the distances of each $C^{\alpha}$ atom from the reconstructed helix
axis was $0.16 \AA$.

The interaxial packing angle between two helices was computed
bewteen the local axis pair for which the minimum
distance of closest approach was achieved. In case the segment of
closest approach (SCA) intersected an inner hinge of the broken
global axis, the local axis direction was defined as the average
of the two corresponding local vectors. The whole discussion concerning
the perpendicularity of both local axis direction with the SCA
applies only to cases involving a terminal local axis, since
face-to-face packing is anyway ensured in case of contact between
inner local axes.
Packing angles are positive if the background helix is rotated clockwise
with respect to the frontal helix when facing them.
The angle $\Omega = 0^{\circ}$ ($\Omega = \pm180^{\circ}$)
corresponds to parallel (antiparallel) helices with respect
to their sequence direction.

Imposing the further requirement that the SCA intersected both local
axis directions at a perpendicular angle within a threshold
$\tau_{\rm max}$ plays a critical role, as discussed in the text.
For the three different $\tau_{\rm max}$ values considered in this work,
$\tau_{\rm max}=7^{\circ},11^{\circ},15^{\circ}$, we collected datasets
of $765$, $837$, $899$ closely packed helix pairs, respectively.

As a final remark we report the mean value of the angle $\Delta\Omega$
which is formed
between consecutive local directions averaged over all inner hinges of all
helical axes which were reconstructed using the procedure discussed above.
We found $\overline{\Delta\Omega}=6.7^{\circ} \pm 4.0^{\circ}$, strongly supporting
the necessity of allowing a similar threshold when imposing angular
constraints as in condition $C2$.

\section*{Results and Discussion}

\subsection*{Geometrical analysis}

We will now sketch the geometrical consequences of relaxing condition
$C2$ within a given threshold, in order to understand by means of a
simple argument how this effects the random reference distribution.  A
comprehensive analytical treatment is in principle feasible but quite
cumbersome. Numerical simulation will ultimately be the preferential
approach to extract the corrected random reference distribution for
interaxial angles.

Within the admitted threshold $\tau_{\rm max}$, cases similar to the
one described in Fig. 1b may now occur. The SCA
between the two helices, which intersects helix 1
at one of its axis ends and helix 2 at some internal point,
forms an angle $\theta_1$ which deviates from $\frac{\pi}{2}$
by less than $\tau_{\rm max}$, whereas the GSCA
does not intersect both helices
(in this situation $\tau_1 \neq0,\tau_2=0$).
If $d < d_c$, the two helices are considered to be in contact.
The condition $\tau_1<\tau_{\rm max}$ is equivalent to the condition
that GSCA intersects helix 1
within a distance $\Delta l_{\rm max}$ from the end of its axis.
This limit is reached when
$\theta_1=\theta_{\rm min}\equiv\frac{\pi}{2}-\tau_{\rm max}$.

There are also cases in which $\tau_1=0$ and $\tau_2 \neq 0$, when the
SCA
intersects helix 2 at one of its axis ends
and helix 1 at some internal point, and cases in which both $\tau_1$
and $\tau_2$ are not zero, and the SCA
intersects both helices at one of their axes ends (Fig. 1c).
In this last case, the geometrical condition implied by $\tau_1+\tau_2
<\tau_{\rm max}$, and involving the distances $\Delta l_1$, $\Delta l_2$
from the end of both helix axes to their intersection with the GSCA,
is more complicated (and is discussed in the caption of Fig. 1).

Using the visual representation of Fig. 2b, we can say that allowing
condition $C2$ to be satisfied within the threshold
$\tau\equiv\tau_1+\tau_2<\tau_{\rm max}$, opens up the possibility for
the starting point of helix 2 to lie in the portion of space (${\cal G}$)
formed by the four lightly shaded parallelograms (two parallel to helix 1
with sides $h_1$ and $\Delta\equiv{\Delta l_{\rm max}} \sin{\Omega}$, and
two parallel to helix 2 with sides $h_2$ and $\Delta$)
and by the four remaining white corner regions (
whose boundaries are defined by hyperboles,
see Figure 2 caption for details), which surround
parallelogram $A$.

To compute the area of ${\cal G}$ is not trivial, because of the
hyperbole-shaped regions, but the result is obviously $\Omega$ dependent.
For example a simple trigonometric calculation shows that:

\begin{equation}
\Delta =
\left\{
\begin{array}{cl}
\frac{d \sin{\tau_{\rm max}}}{
\sqrt{\sin^2{\Omega}-\sin^2\tau_{\rm max}} }  &
\left|\sin{\Omega}\right| > \sin{\tau_{\rm max}} \\
\infty &  \left|\sin{\Omega}\right| \leq \sin{\tau_{\rm max}}
\end{array}
\right.
\end{equation}

This implies that the area of ${\cal G}$ diverges for
$\left|\sin{\Omega}\right| \leq \sin{\tau_{\rm max}}$,
which already points towards the fact that relevant corrections
are possible even for small values of $\tau_{\rm max}$.

To conclude our analysis we need to enforce also condition $C1$. By
using the pythagorean theorem it is easy to see that for any fixed
$d<d_c$, $C1$ is satisfied for any point (thought as starting point of
helix 2) whose distance $\Delta s(d_R) = \sqrt{d_R^2-d^2}$ from
parallelogram $A$ is less than $\Gamma =\Delta s(d_c)=
\sqrt{d_c^2-d^2}$. These points belong to the region $\cal{G}'$, shown
in Fig. 2c, formed by the 4 parallelograms (two of sides $h_1$ and
$\Gamma$ and two of sides $h_2$ and $\Gamma$) and the four circular
sectors (of radius $\Gamma$) surrounding A.  The area of ${\cal
G}^{'}$ is independent on $\Omega$ (see caption of Fig. 2c).

All the points in the portion of space ${\cal H} = \cal{G} \cap
\cal{G}^{'}$ resulting from the intersection of
${\cal G}$ and ${\cal G}^{'}$
satisfy  both condition $C1$ and $C2$.

Therefore, the probability $P^{\tau_{\rm max}}(d)$ of selecting
a particular angle $\Omega$ with a tolerance $\tau_{\rm max}$, at a
given  distance $d < d_c$, is given by the product of the spherical
polar term $\sin{\Omega}$ \cite{Bowie97} and a term proportional to the
area of parallelogram $A$ {\em plus} the area of ${\cal H}$.

When $|\sin\Omega| < \sin{\tau_{\rm max}}$, Eq.2 shows that  $\Delta$
 diverges and thus:
\begin{equation}
P^{\tau_{\rm max}}(d) \sim \sin{\Omega} \left( h_1 h_2 \sin{\Omega} +
2 (h_1+h_2) \Gamma + B \right)
\end{equation}
where $B$ is the area of the corner regions in ${\cal H}$.

Since $\Gamma$ does not depend on $\Omega$, the second term in the
bracket will eventually dominate the small $|\sin\Omega|$ behaviour of
$P^{\tau_{\rm max}}(d)$. In practice, the actual relevance of this
effect can be appreciated only after integrating $P^{\tau_{\rm
max}}(d)$ over $d$, since the relative weights of the different terms
in the bracket vary with $d$.  Such computation is quite cumbersome
and we have rather chosen to get $P^{\tau_{\rm max}}$ by simulating
random helices which satisfy the contact conditions within a threshold
$\tau_{\rm max}$, and to extract numerically the normalized
histograms.

\subsection*{Numerical simulations}

In order to compute the reference distribution
$P^{\tau_{\rm max}}\left(\Omega\right)$ for interaxial packing angles we
generated {\em random} helix pairs by means of computer simulations,
and then selected them with the same conditions, $C1$ and $C2$,
used in extracting histograms from real helices in native protein
structures. Note that when computing the reference distribution we
did not take steric effects into account; in other words
random helices might overlap.

More specifically, we constructed ideal discretized helices with the
same geometrical properties of $\alpha$-helices in real proteins,
i.e. twist per residue $99.1^{\circ}$, rise per residue $1.45 \AA$,
radius $2.3 \AA$ (as is the case when considering $C_{\alpha}$ atoms).
We chose to generate helices consisting of $11$ residues, the average
length in the dataset of real helices that we collected from the PDB.
Keeping fixed the position of the first helix, the second helix
was placed by firstly choosing randomly the midpoint of its axis
within a sphere of radius $15 \AA$ centred in the
midpoint of the first helix axis, and then selecting, again randomly,
both the direction of its axis and the twist of its first residue.

Boundary effects might be relevant, when the radius of the sphere in
which the second helix is generated is too small with respect to the
helix length and helix finiteness is effectively reduced.
We made sure that our results did not change when increasing the
radius of the sphere in which the second helix axis is placed.

We generated $5\cdot10^7$ random helix pairs, $29915472$ out of which
satisfied condition $C1$ of having at least one pair of residues
distant less then $5.5 \AA$. Condition $C2$ was then applied, in the
same way as explained in Methods for the real helices data set, with the
whole axis being now a segment in the ideal case.  In this way we
generated $11467456$, $13033815$, $14553626$ helix pairs,
respectively, for the three different $\tau_{\rm max}$ values
considered in this work, $\tau_{\rm
max}=7^{\circ},11^{\circ},15^{\circ}$.

In Fig. 3 we plot the corresponding random reference distributions
$P^{\tau_{\rm max}}\left(\Omega\right)$, comparing them with the ideal
($\tau_{\rm max}=0$) case: $P(\Omega)\sim\sin^2{\Omega}$. The
difference, although due to a very subtle effect is substantial, and
it is clearly seen that a new regime is present for
$|\sin\Omega|<\sin\tau_{\rm max}$, as expected from the previous
discussion.

\subsection*{Reanalysing experimental results}

To test the effect of this new reference distribution we have
recomputed an experimental distribution of interaxial angles (see
Methods for details) by analysing a databank of 600 proteins and using
a contact threshold $d_c=5.5 \AA$ and a tolerance $\tau_{\rm
max}=7^{\circ}$, corresponding the intrinsic uncertainty
$\overline{\Delta\Omega}=6.7^{\circ}$ estimated when determining local
axis directions (see Methods).  The histogram is reported in the upper
panel of Fig. 4: in order to obtain good statistics for each single
bin, we gathered results every $15^{\circ}$. The histogram is
consistent with previous analysis \cite{Walther1,Bowie97,Walther2}.
In the middle panel of Fig. 4 we plot $P^{7^{\circ}}(\Omega)$ and, for
reference, $P(\Omega)$ normalized with the same binning as in the
previous panel. Finally, in the lower panel of Fig. 4 we present the
histogram both rescaled with $P^{7^{\circ}}(\Omega)$ and with
$P(\Omega)$. The results are clear and striking: the correct
distribution $P^{7^{\circ}}(\Omega)$ removes the spurious peaks at
$0^{\circ}$ and $\pm 180^{\circ}$, whereas $4$ out of the $6$
predicted packing angles \cite{Walther1} arise as clean local maxima
in the expected positions. Only two of the predicted packing angles
are clearly not favoured according to our analysis.  Remarkably, the
two peaks theoretically predicted at $143.9^{\circ}$ and
$-158^{\circ}$, that were seen as shoulders in the unrescaled
histogram, are instead placed in the correct position (within binning
uncertainty) in the rescaled histogram.  We also note that a residual
preference for parallel and antiparallel alignment of two contacting
helices cannot be ruled out even after proper rescaling. Because of
small statistics, it is difficult to ascertain whether this is a
genuine effect or is just due to the tail superposition of different
preferential angle peaks.  In order to further test both the
robustness of the results and the full procedure we have obtained
statistical histograms from the experimental data using three
different thresholds, $\tau_{\rm max}$.  Since we can compare them
with the corresponding random distribution we should expect three
similar rescaled histograms. This is nicely confirmed by Fig. 5, where
the three rescaled histograms show a very good overlap, within the
statistical uncertainty. We also note that generally, the greater the
threshold used to relax condition $C2$, the more blurred the peaks
corresponding to the sterically preferred angles. This confirms that
condition $C2$ is needed to ensure face-to-face packing and the
applicability of steric models.

\section*{Conclusion}

In this paper we have shown that the calculation of the probability
distribution of interaxial angles between random finite helices
which are in contact is not a trivial  geometric problem,
because of the approximations introduced to ensure face-to-face
packing between contacting
helices. Such approximations are unavoidable, due to the
imperfect shape of natural occurring helices which do not have  well
defined axes. Although analytical results can be found to estimate
the correct random distribution, the simplest way to obtain it
consists in using numerical simulations.  We have presented a
re-analysis of the distribution of packing angles rescaled with our
new reference distribution and we have  found  a remarkable
agreement with the packing angles predicted by steric
models \cite{Walther1}.



\section*{Acknowledgments}

We are indebted to Amos Maritan for stimulating discussions and advice,
Giuseppe Zanotti  for helpful suggestions and
Iksoo Chang for his invaluable help in
providing us with the protein databank used in ref. \cite{Iksoo}.
We thank Harvey Dobbs for a critical reading of the manuscript.
This work was supported by
INFM, MIUR COFIN-2001 and FISR 2002.
\\


\section*{Figure Legends}

\subsection*{Figure 1}

a) The global segment of closest approach (GSCA), of length $d$,
between two straight lines is,
by definition, perpendicular to both of them. The
segment of closest approach (SCA), of length $d_r$, between two
helices (schematically represented by cylinders) surely coincides with
the global one if it intersects both helices within their axis
length. This fact is always guaranteed if the two helices are assumed
to have infinite length (hypothesis of Ref. \cite{Bowie97}).

b) The two finite helices have a SCA that is not coincident with the
GCSA. The latter intersects helix 1 at a distance $\Delta l_1$
from its end and may (as in the picture) or may not intersect helix 2.
The former joins the end point of helix 1 to a point inside helix
2. The angle formed with helix 1 ($\theta_1$) is deviating from
$\pi/2$ by an amount $\tau_1$. Instead $\theta_2=\frac{\pi}{2}$ and
$\tau_2=0$.  The trigonometric relation expressing $\Delta l_1$ as an
increasing function of $\tau_1$ can be easily obtained:
$
\Delta l_1(\tau) = \frac{d \sin{\tau_1}}{\sin{\Omega}
\sqrt{\sin^2{\Omega}-\sin^2\tau_1} }
$.
When $\tau_1 < \tau_{\rm max}$, condition $C2$, defined in the text, is
still fulfilled.  This is equivalent to require that
$\Delta l_1 \leq \Delta l_{\rm max}\equiv \Delta l_1(\tau_{\rm max})$.
Notice that $\Delta l_1(\tau_1)$ diverges
when $\tau_1$ approaches $\Omega$ from below, and consequently, when
$\left|\sin\Omega\right| < \sin\tau_{\rm max}$, $\Delta l_{\rm max}$ diverges. In such
regime the condition C2 (but obviously not C1 because $d_r$ may also
diverge) is always satisfied within the allowed tolerance.
A similar situation and a similar analysis can be done for
$\tau_1=0$ and $\tau_2 \neq 0$.

c) GSCA is neither
intersecting helix 1 nor helix 2 and the SCA
is joining two endpoints. In the case
depicted in the figure the two helices are placed in opposite sides
with respect to the GSCA. The threshold condition
$\tau_1+\tau_2<\tau_{\rm max}$ may be translated into the following one
involving $\Delta l_1$ and $\Delta l_2$: $((\Delta l_1+\Delta l_2\cos\Omega)
(\Delta l_2+\Delta l_1\cos\Omega)-\sqrt{(d^2+\Delta l_1^2\sin^2\Omega)
(d^2+\Delta l_2^2\sin^2\Omega)})$ $/$ $((\Delta l_1+\Delta l_2\cos\Omega)
\sqrt{d^2+\Delta l_1^2\sin^2\Omega}+(\Delta l_2+\Delta l_1\cos\Omega)
\sqrt{d^2+\Delta l_2^2\sin^2\Omega}) >\cot\tau_{\rm max}$.
Situations in which the two helices are placed on the same side
with respect to the GSCA are also possible. In this case the
condition becomes $((\Delta l_1-\Delta l_2\cos\Omega)
(\Delta l_2-\Delta l_1\cos\Omega)-\sqrt{(d^2+\Delta l_1^2\sin^2\Omega)
(d^2+\Delta l_2^2\sin^2\Omega)})$ $/$ $((\Delta l_1-\Delta l_2\cos\Omega)
\sqrt{d^2+\Delta l_1^2\sin^2\Omega}+(\Delta l_2-\Delta l_1\cos\Omega)
\sqrt{d^2+\Delta l_2^2\sin^2\Omega}) >\cot\tau_{\rm max}$.

\subsection*{Figure 2}

a) In this figure, the position of helix 1
(represented by a vector of length $h_1$)
and a vector orthogonal to the paper plane, determining the
direction of the GSCA to a second helix, are fixed.
Given a particular packing angle
$\Omega$, in order
to satisfy condition C2 without any tolerance ${\tau_{\rm max}}$,
a second finite helix axis (of length $h_2$) may then be
placed only within a plane parallel to the paper plane and such that
its starting point
is inside the dark shaded parallelogram $A$,
Otherwise the SCA between the two
finite axes would no longer be perpendicular to both of them.
The area of parallelogram $A$ is $h_1 h_2 \sin(\Omega)$.

b) The tolerance $\tau_{\rm max}$
with which the condition of perpendicularity between the SCA
and the two helices is implemented enlarges the portion of space
where the starting point of helix 2 can be placed. The region $\cal{G}$
formed by the four lightly shaded
parallelograms and the four  white corner  regions, which together
surround parallelogram $A$, can now be exploited.
The two parallelograms parallel to helix 2,
whose sides are $h_2$ and $\Delta=\Delta l_{\rm max}\sin{\Omega}$,
are due to the geometric arrangement described in Fig. 1b
with $\tau_1 \neq 0$ and $\tau_2=0$ (the vector with
the dashed starting point represent the possible location of helix 2 in such context).
The two other parallelograms, parallel to helix 1,
with sides $h_1$ and $\Delta$, correspond to the similar case
$\tau_1 = 0$ and $\tau_2 \neq 0$.
The four remaining white regions, corresponding to the case
$\tau_1 \neq 0$ and $\tau_2\neq 0$, are all limited by hyperbola-like curves.
The two of them subtending an obtuse angle are due to the
geometric arrangement described in Fig. 1c, corresponding to
the two helices being placed in opposite sides
with respect to the GSCA  (the vector with
the white starting point represent the possible location of helix 2 in such context).
The two of them subtending an acute angle correspond to the two helices
being placed in the same side with respect to the GSCA.
In both cases it is possible to find the equation describing such
boundary curves, which we omit here for the sake of simplicity.
The area of $\cal{G}$ is $2(h_1+h_2)\Delta$ plus a non trivial
contribution coming from the four corner regions.

c) The distance constraint C1 limits by itself the portion of
plane surrounding parallelogram $A$ where the starting point of
the second helix can be placed. This region $\cal{G}^{'}$
is simply the locus of
points whose Euclidean distance from $A$ in the paper plane is less than
$\Gamma\equiv\sqrt{d_c^2-d^2}$.
For given values of $d$ and  the distance threshold $d_c$, this region
is thus formed by the four lightly shaded parallelograms and by the
four white circular sectors. The area of $\cal{G}^{'}$ is $2(h_1+h_2)\Gamma
+\pi\Gamma^2$.

\subsection*{Figure 3}

Ideal `random' distribution of interaxial packing angles
for four different values of the angle threshold $\tau_{\rm max}$
used in applying condition $C2$; $\tau_{\rm max}=0^{\circ}$ (upper left panel),
$\tau_{\rm max}=7^{\circ}$ (upper right panel),
$\tau_{\rm max}=11^{\circ}$ (lower left panel),
$\tau_{\rm max}=15^{\circ}$ (lower right panel).
The four distributions were computed by means of numerical
simulations described in Methods. Since ideal helices do not have
a preferential order, the histograms show in this figure have been
restricted to the $-90^{\circ}<\Omega<90^{\circ}$ region.
The $14553626$ packing angle values collected in
the $\tau_{\rm max}=15^{\circ}$ histogram were then successively
filtered out by the more and more restrictive
$\tau_{\rm max}=11^{\circ}$, $\tau_{\rm max}=7^{\circ}$,
$\tau_{\rm max}=0^{\circ}$ threshold conditions to generate the
$13033815$, $11467456$, $8646994$ data, respectively, collected in the
corresponding histograms.
The dashed line is the fit of the $\tau_{\rm max}=0^{\circ}$ data,
obtained enforcing condition $C2$ in a strict way,
to the expected $P(\Omega)\sim\sin^2\Omega$ distribution,
which is then reported for comparison in all other histograms.
All histograms were constructed with a bin width of $1.5^{\circ}$.
Histograms are normalized in such a way that a flat distribution would
correspond to a constant $1$ height.

\subsection*{Figure 4}

Upper Panel. Experimental `bare' unrescaled distribution of interaxial packing
angles between contacting helices. Data were obtained with a distance threshold
$d=5.5\AA$ for interhelical $C_{\alpha}$ atom pairs and an angular threshold
$\tau_{\rm max}=7^{\circ}$ for condition $C2$.
The histogram is normalized in such a way that a flat distribution
would correspond to a constant $1$ height.

Middle Panel. Ideal `random' distribution which have to be used to rescale
the `bare' experimental distribution, in order to reveal true packing angle
preferences. The two different distributions obtained with either 
$\tau_{\rm max}=0^{\circ}$ (thick line) or
$\tau_{\rm max}=7^{\circ}$ (filled columns) are shown.
Histograms are normalized in such a way that a flat distribution would
correspond to a constant $1$ height.

Lower Panel. Rescaled distribution of interaxial packing angles.
The experimental `bare' distribution in the upper panel is divided by the
ideal `random' one in the middle panel, obtained with either
$\tau_{\rm max}=0^{\circ}$ (blue line) or
$\tau_{\rm max}=7^{\circ}$ (filled columns).
Histogram heights greater than $1$ correspond to preferential packing angles,
whereas heights lower than $1$ correspond to disfavoured packing angles.

All histograms were constructed with a bin width of $15^{\circ}$.
Arrows in the upper and lower panels mark the values of the six preferred
packing angles predicted by Walther {\it et al.}\/ 
\cite{Walther1}; $\Omega_{\rm abc}=-37.1^{\circ},-97.4^{\circ},22.0^{\circ}$,
each represented twice with a periodicity of $180^{\circ}$.

\subsection*{Figure 5}

Rescaled distribution of interaxial packing angles between contacting
helices for three different values of the angle threshold $\tau_{\rm max}$
used in enforcing condition $C2$;
$\tau_{\rm max}=7^{\circ}$ (light grey filled columns),
$\tau_{\rm max}=11^{\circ}$ (dark grey line),
$\tau_{\rm max}=15^{\circ}$ (black line).
Each histogram was obtained by dividing the experimental unrescaled
distribution by the corresponding ideal reference one
(we used the three distributions represented in Fig. 3 with a
different binning).
All histograms were constructed with a bin width of $15^{\circ}$.
Note that the histograms are not normalized, being computed as the
ratio of two normalized histograms.
The arrows mark the values of the the six optimal packing angles predicted by
Walther {\it et al.}\/ \cite{Walther1}.

\newpage

\begin{figure}
\centering
\epsfig{width=9.0cm,angle=-90,file=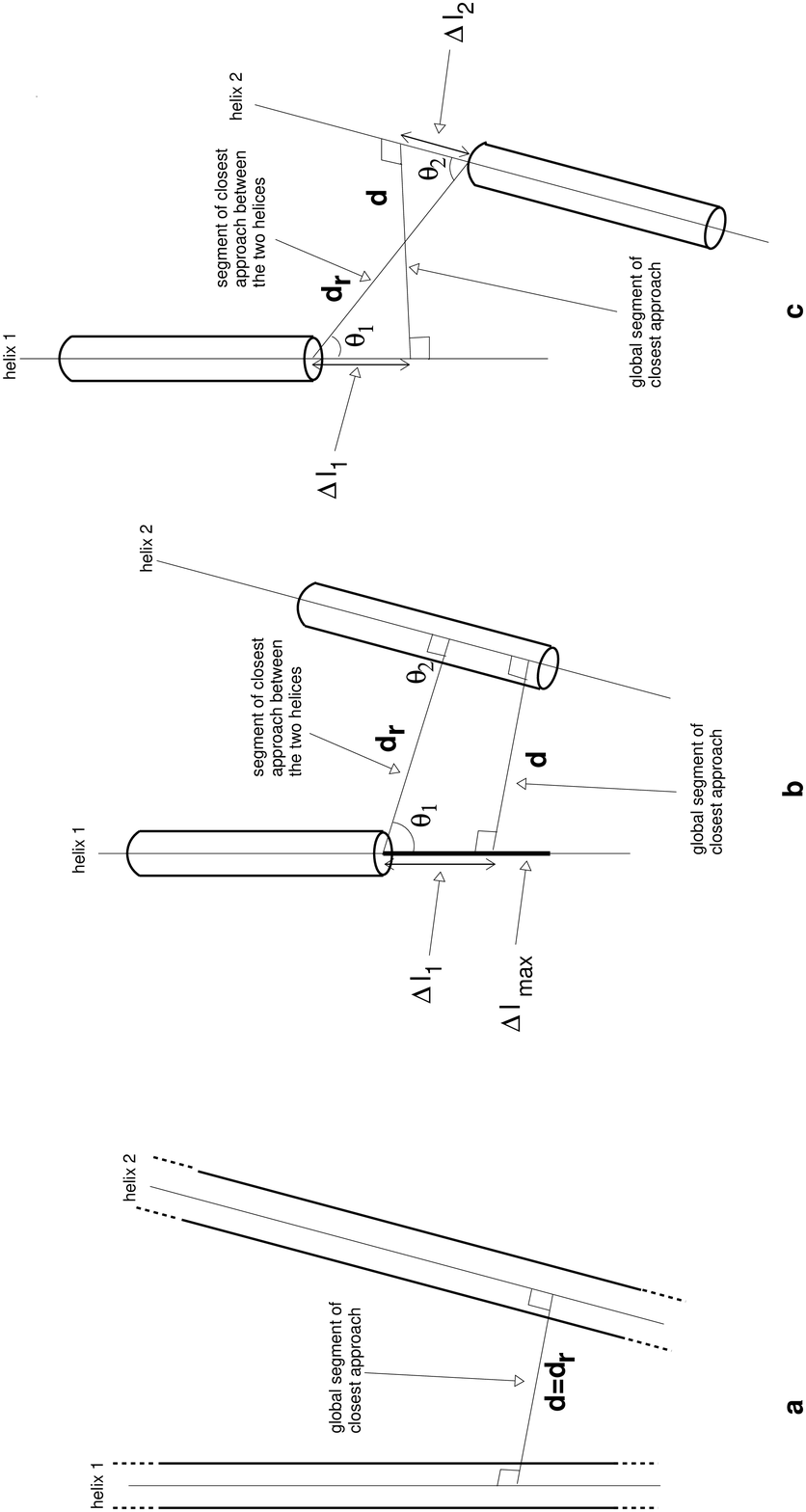}
\end{figure}

\begin{center}
{\bf Figure 1}
\end{center}

\newpage

\begin{figure}
\centering
\epsfig{width=9.0cm,angle=-90,file=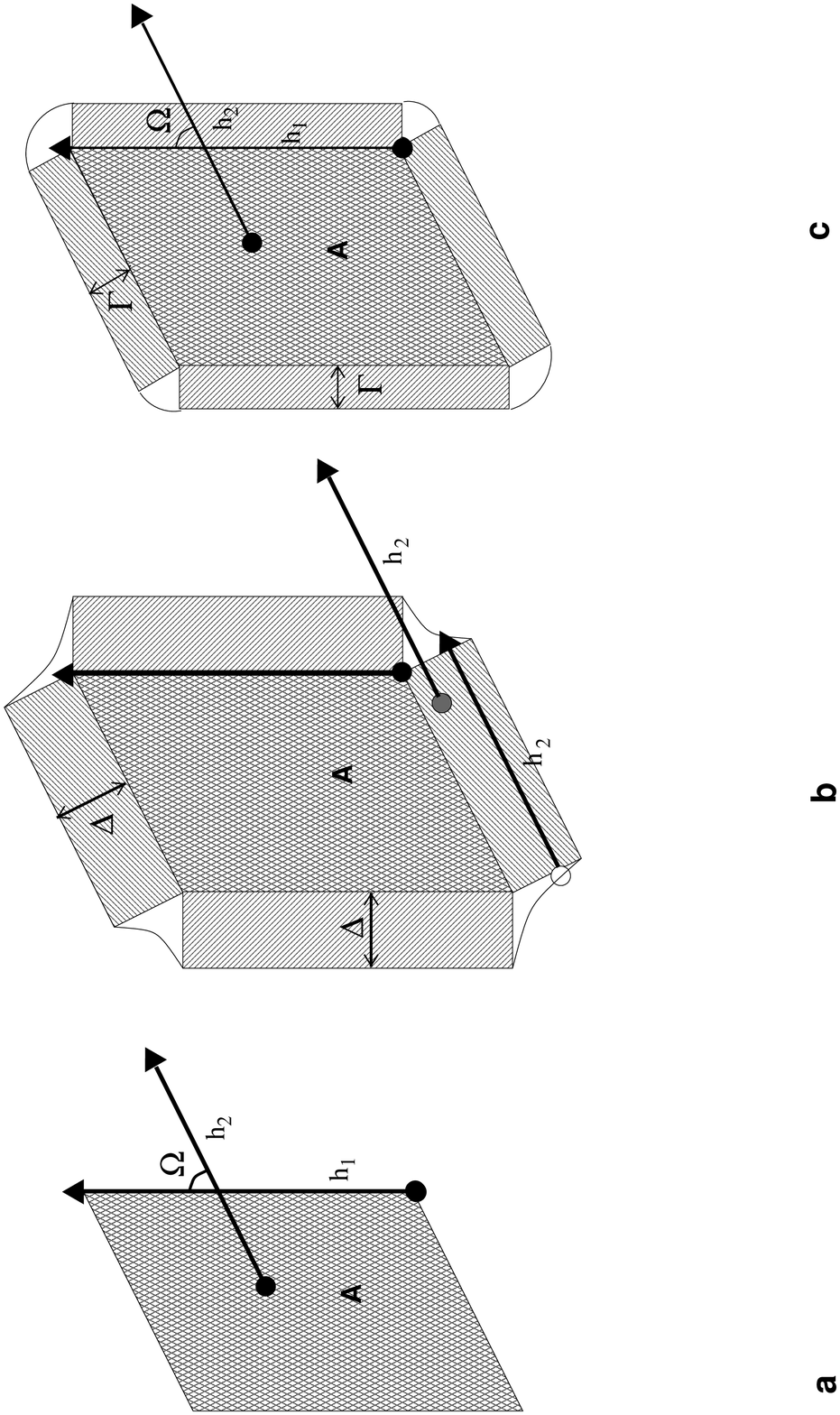}
\end{figure}

\begin{center}
{\bf Figure 2}
\end{center}

\newpage

\begin{figure}
\centering
\epsfig{height=8.0cm,file=fig3.ps}
\end{figure}

\begin{center}
{\bf Figure 3}
\end{center}

\newpage

\begin{figure}
\centering
\epsfig{height=8.0cm,file=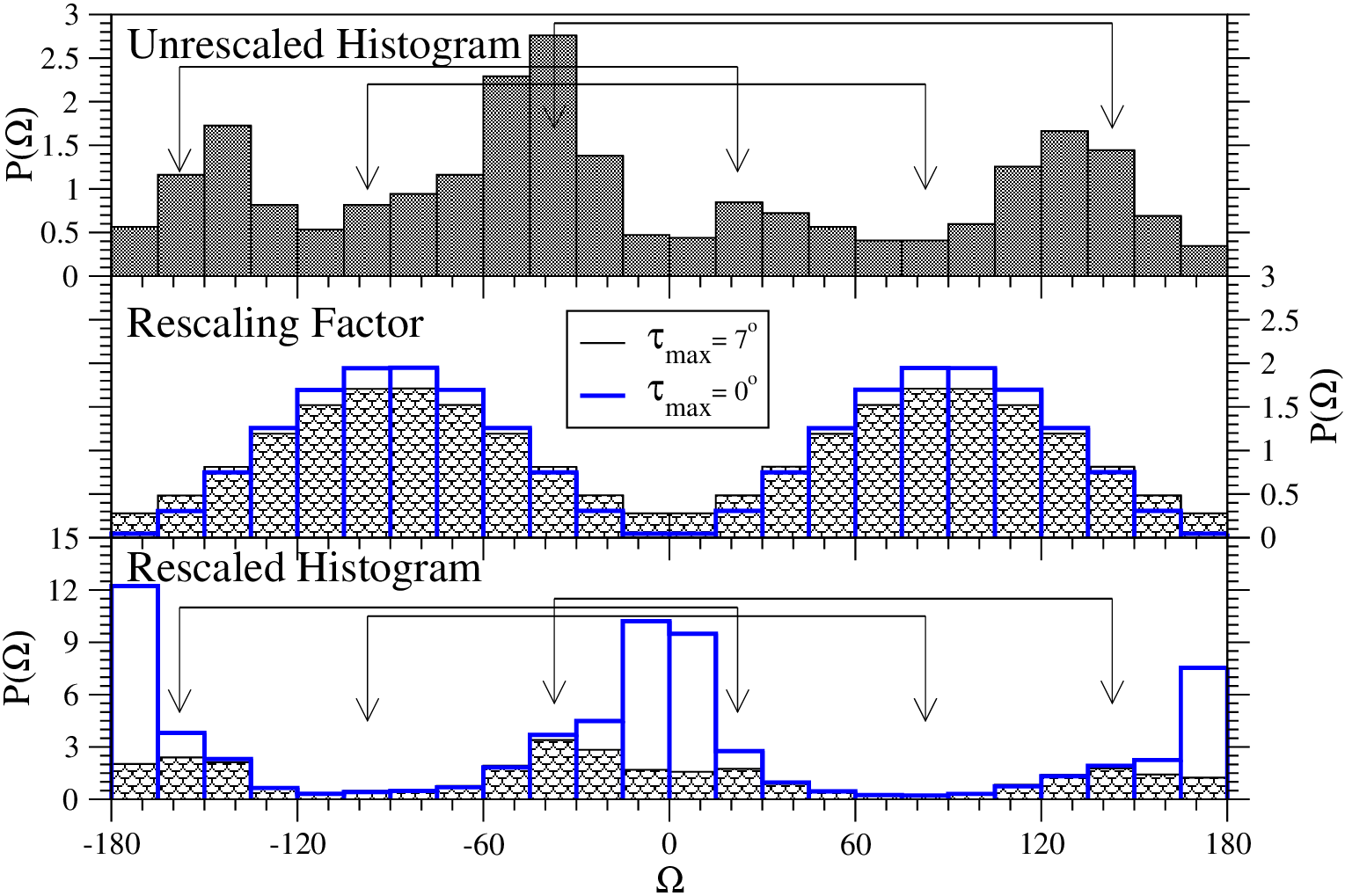}
\end{figure}

\begin{center}
{\bf Figure 4}
\end{center}

\newpage

\begin{figure}
\centering
\epsfig{height=6.5cm,file=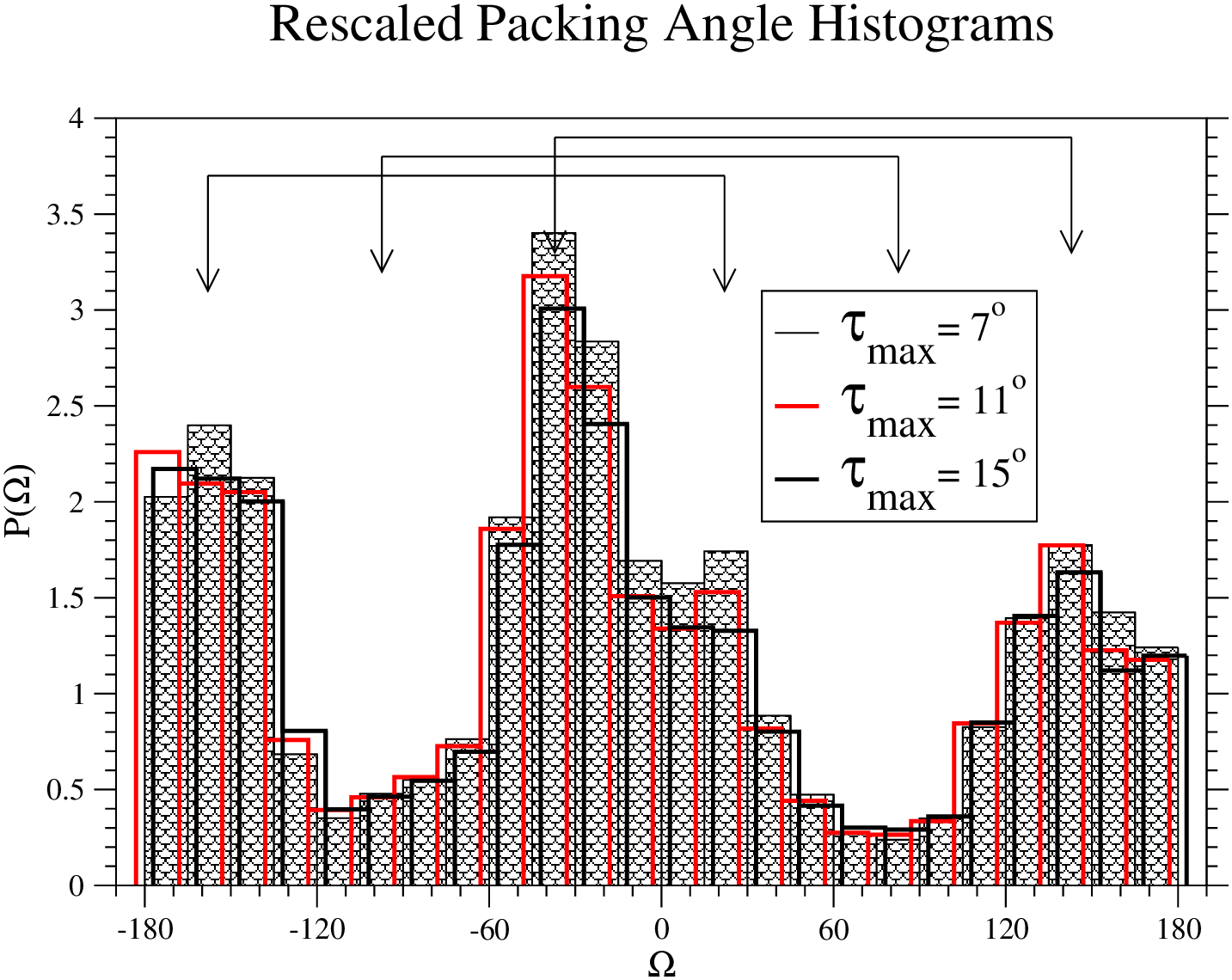}
\end{figure}

\begin{center}
{\bf Figure 5}
\end{center}

\end{document}